\documentclass[prd,showpacs,twocolumn,amsmath,amssymb]{revtex4}
\usepackage{amssymb}
\usepackage{mathrsfs}
\usepackage{txfonts}

\usepackage{graphicx}
\usepackage{dcolumn}
\usepackage{bm}

\begin{document}

\title{The entanglement of two qubits coupled ultrastrongly with a quantum oscillator}
\author{Guihua Tian}\email{tgh-2000@263.net, tgh20080827@gmail.com }
 \affiliation{School of Science, Beijing
University of Posts And Telecommunications. Beijing 100876 China.}
\author{Shuquan Zhong}\email{shuqzhong@gmail.com}
 \affiliation{State Key Laboratory of Information Photonics and Optical
Communications, \\ Beijing University of Posts And
Telecommunications.
 Beijing 100876 China.}
\begin{abstract}
We explore the possibility of the controlled manipulation of the entanglement of two qubits with an external apparatus, the Rabi
Hamiltonian. The novel results show that the initially entangled two qubits could have very high probability to stay unchanged by the
coherent state of the photon with chosen parameters. If the inter-qubit coupling strength is negative, their entanglement could
be kept for a much longer time by a relatively strong external quantum control with the suitable and novel choice of some
parameters. Furthermore, their entanglement will not show any sudden death and revival. All these results are different from the previous
studies and they are given with reasonable explanations. The maintaining of the entanglement of the two qubits with very high
fidelity will be very helpful for the quantum information process.
\end{abstract}
\maketitle

Entanglement is at the heart of quantum process. Entangled states could be used to manipulate many interesting tasks, like entanglement purification, factorization of integers, random searches, etc. \cite{van,shor,ekert}. It is crucial for the controlled manipulation of the entanglement with an external apparatus. However, this is still proven extremely challenging.

In the letter, we will focus on the preservation of the entanglement of two qubits, which is entangled initially. We also take the control apparatus as a continuous variable (CV) interaction, whose Hamiltonian is that of Rabi model\cite{rabi}. Our approach differs from that of previous studies in that both the  control field and coupling of qubits and quantum field are strong.

Theoretically, there are investigations of entanglement generation and maintaining of two qubits by use of Jaynes-Cummings (JC) model
Hamiltonian, where initially weakly squeezed states of CV transfers their entanglement to the two qubits\cite{yu}. Further extended
study shows the field entanglement in JC cavity can be transferred to a pair of atoms, and reversed transfer of
entanglement\cite{krau,sorn,kim,kim2,zhou}. Later, coherent-state control of a pair of non-local atom-atom entanglement between two spatially
separated sites was also studied \cite{m}. These results also reveal some time-dependent entanglement death and rebirth effects.

JC model is an approximation of the Rabi Hamiltonian under the condition of  weak coupling of qubit and photon (quantum oscillator)
and their resonance, which result in rejecting the non-energy-conserving term (or the count-rotating-wave terms)\cite{mand}. The strong and ultra-strong coupling region of qubit and oscillator is now feasible in circuit QED recently\cite{12011364,Joo,stef,koch,cres,dica,jing,fice,fili,iris,iris2,agar,qing,asha}, and it has triggered renewed and extensive research interest in Rabi model, These studies provide many new and anti-intuitive results for the Rabi model\cite{iris,iris2,agar,qing,asha}. The strong coupling and zero detuning is treated in Ref.\cite{sorn}. Further, ultra-strong coupling and large detuning is treated and interesting effects appear, including frequency modification, collapse and revival of Rabi oscillation for one qubit with the initial state of the oscillator being  thermal or coherent\cite{iris,iris2,agar,qing,asha}. Later, the two qubits' entanglement is also studied with the death and revival phenomena for the initial coherent state of the oscillator \cite{agar}.

It is generally believed that the control field initially  in its coherent state will result in death and rebirth (or revival) of the
entanglement for the two initial entangled qubits, which has been shown in Refs\cite{m,agar} , either in weak coupling model
of JC Hamiltonian or in the strong or ultra-strong model of Rabi Hamiltonian.

In Ref.\cite{agar}, the entanglement of the two qubits coupled strongly with a quantum oscillator is studied under the conditions that the coupling stength  $|\beta|\le 0.25$,  $|\alpha|\gg 1$ and  $|\alpha\beta|\ll 1$. (the parameter $\alpha$ is related with the everage number $N$ of  the photon's coherent state by $N=|\alpha|^2$. These conditions are very restrict. In the letter, we donot require the two conditions  $|\alpha|\gg 1$ and  $|\alpha\beta|\ll 1$, and permit the two qubits have intra-qubit interaction. Then we restudy the problem obtain some novel results. Our results show  that the initially entangled two qubits can have very high probability to stay unchanged by the coherent state of the photon with chosen parameters. If the inter-qubit couples with each other with negative coupling strength, their entanglement can be kept for a long time by a relatively strong external quantum control with the suitable and novel choice of some parameters and there will be no any sudden death and revival phenomena at all. We will provide an explanation for the results in later.

Some nots to be added. Super-conducting circuits (SC) are promising as quantum qubits for quantum process because they can be flexibly  designed with controllable  parameters. Some SCs, like the phase qubit, the capacitively shunted flux qubit and the transom qubit, are immune to the charge noise. These qubits are the focus of our investigation. So,  we do not consider the effect of the environment noise on the qubits any longer in the following\cite{12011364,Joo,stef,koch}.

The Hamiltonian of the extended Rabi model  is given by
\begin{eqnarray}\label{H for qutrit}
H&=&\hbar\omega
a^{\dagger}a+\hbar\omega(\sigma_z^{(1)}+\sigma_z^{(2)})(\beta a+\beta^*a^{\dagger})
\nonumber\\ && -\frac12\hbar\omega_0(\sigma_x^{(1)}+\sigma_x^{(2)})-\kappa\hbar\omega_0\sigma_x^{(1)}\sigma_x^{(2)},
\end{eqnarray}
where $\sigma_x^{(i)},\ \sigma_z^{(i)}, \ i=1,2$ are the ith qubit's Pauli matrices and $a^{\dagger},\ a$ are the creation and
annihilation operators of the quantum oscillator. In the ultrastrong coupling regime with large detuning $\hbar\omega\gg  \hbar\omega_0$,
the free part of the Hamiltonian is
\begin{eqnarray}\label{H0 forqutrit}
H_0=\hbar\omega
a^{\dagger}a+\hbar\omega(\sigma_z^{(1)}+\sigma_z^{(2)})(\beta a+\beta^*a^{\dagger}).
\end{eqnarray}
It means the quantum oscillator of the Rabi system is focused first and it is influenced by its interaction with the two qubits. Suppose
$\sigma_z|\uparrow \rangle=|\uparrow \rangle,\ \sigma_z|\downarrow \rangle=-|\downarrow \rangle$, then the composite states for the two
qubits are
\begin{eqnarray}
|1,1\rangle&=&|\uparrow \uparrow\rangle, \ |1,-1\rangle=|\downarrow\downarrow\rangle
\nonumber\\|1,0\rangle
&=&\frac1{\sqrt2}(|\uparrow\downarrow\rangle+|\downarrow\uparrow\rangle,
\nonumber\\
|0,0\rangle
&=&\frac1{\sqrt2}(\uparrow\downarrow\rangle-|\downarrow\uparrow\rangle,\label{22}
\end{eqnarray}
with $|1,0\rangle,\ |0,0\rangle$ being two of the Bell states, the fully entangled states for the two qubits. The energy states for
free Hamiltonian are the displaced number states $|N_m\rangle=D(-m\beta)=\exp{(-m\beta a^{\dagger}+m\beta^*a)}$ with
the corresponding states $|j,m\rangle, j=1,0, m=0,\pm1$ of the two qubits in Eq.(\ref{22}), that is,
\[H_0|j,m\rangle|N_m\rangle=(N^2-m^2\beta^2)\hbar\omega|j,m\rangle|N_m\rangle.\]
More ever, the states $|0,0\rangle|N_0\rangle,\ N=0,1,2,\cdots$ are also the energy-states of the whole Hamiltonian $H$. So the two
qubits will remain fully entangled if the composite system initially stay in one of $|0,0\rangle|N_0\rangle,\ N=0,1,2,\cdots$. In the
following, we will no longer consider these states as they evolve independently.

The coupling of the quantum oscillator with the two qubits makes it become several different oscillators, each with a unique displaced
number states $|N_m\rangle,\ m=0,\pm1,\ N=0,1,2,\cdots$.  Then
\begin{eqnarray}\label{H' for qutrit}
H'=-\frac12\hbar\omega_0(\sigma_x^{(1)}+\sigma_x^{(2)})-\kappa\hbar\omega_0\sigma_x^{(1)}\sigma_x^{(2)}
\end{eqnarray}
serves as a perturbation term. Under the large detuning assumption, the qubits only induce the oscillators to transition between its
differently displaced number states $|N_m\rangle,\ m=0,\pm1$ but with the same number $N$, other kind transition is omitted. This is
the adiabatic approximation (AA)\cite{iris,agar}.

By the method of AA, the  energystates of the whole Hamiltonian $H$ are easy to obtain,
\begin{eqnarray}
  \tilde{E}_{N,0}^0&=& \hbar\omega\left(N-\beta^2-\Omega_{2N}\right)\label{e0}, \\
  |\tilde{E}_{N,0}^0\rangle
&=&\frac1{\sqrt2}(|1,1\rangle|N_1\rangle-|1,-1\rangle|N_{-1}\rangle)\label{e0s}, \\
  \tilde{E}^0_{N,\pm} &=& \hbar\omega\left(N-\frac{\kappa}{\hbar\omega}
  +\frac{\tilde{T}_0\pm \sqrt{\tilde{T}_0^2+8\Omega^2_{1N}}}2\right), \nonumber \\
  \label{epm}  \\
|E^0_{N,\pm}\rangle
&=&\frac1{\tilde{L}_{N,\pm}}(|1,1\rangle|N_1\rangle+|1,-1\rangle|N_{-1}\rangle)
\nonumber\\ && +
 \frac{\tilde{Y}_{N,\pm}}{\tilde{L}_{N,\pm}} |1,0\rangle|N_0\rangle\label{epms},
 \end{eqnarray}
 where
 \begin{eqnarray}
\Omega_{1N} &=& -\frac{1}{\sqrt2}\frac{\omega_0}{\omega}\exp{(-\frac{\beta^2}2)}L_N(\beta^2) ,\\
  \Omega_{2N} &=& -\frac{\kappa}{\hbar\omega}\exp{(-2\beta^2)}L_N(4\beta^2),\\
  \tilde{T}_0 &=&  -\beta^2+\frac{\kappa}{\hbar\omega}+\Omega_{2N}
  \label{t0new},\\
\tilde{Y}_{N,\pm} &=& \left(\frac{-\tilde{T}_0\pm \sqrt{\tilde{T}_0^2+8\Omega^2_{1N}}}{2\Omega_{1N}}\right), \\
  \tilde{L}^2_{N,\pm} &=&  \tilde{Y}^2_{N,\pm} +2,\ N=0,1,2,\cdots.
\end{eqnarray}

It is very important to for the initially entangled two qubits remain entangled for long time. The coupling with the oscillator
will make the qubits partly lose their entanglement easily. The convenient states for the quantum oscillator will be the coherent
and displaced coherent states. It is intuitively felt that that these states of the quantum oscillator will generally result in some
degree of the disentanglement of the two qubits. The following will show the intuition generally is right.

Suppose the two qubits initially fully entangled as \[|1,0\rangle=\frac1{\sqrt2}(|\uparrow,\downarrow\rangle+|\downarrow,\uparrow\rangle)\]
and the quantum oscillator is in its coherent state $|\alpha\rangle$, then the probability for they evolve into other states $|1,1\rangle=|\uparrow,\uparrow\rangle$ or $|1,-1\rangle=|\downarrow,\downarrow\rangle$ is
\begin{eqnarray}
T(t)&=&\sum_{N=0}^{\infty}p(N)\frac{\tilde{Y}_{N,+}^2}{\tilde{L}_{N,+}^4}\bigg(1-\cos
(\omega_{N,0}t)\bigg),\label{t0}
\end{eqnarray}
where $p(N)$ is the probability of $N$ photons in the coherent state $|\alpha\rangle$
\[p(N)=\frac{e^{-\alpha*\alpha^*}|\alpha|^{2N}}{N!}.\]
Fig.(\ref{fig1}) are numerical results of several groups of different parameters $\frac{\omega_0}{\omega},\
\frac{\kappa}{\hbar\omega},\ \beta,\ |\alpha|^2$.
Generally, $T(t)$ will be non-zero, as Figs.(\ref{fig1})-(\ref{fig2}) show. The qubits stay in in their initial state  with the probability $1-2T(t)$. If $T(t)$ can not be neglected, then the qubits can not stay in in their initial state completely and the entanglement between them
decreases.

In the strong coupling regime of qubits and oscillator, the dynamics of the system differs greatly from that of JC model due to the
complex relations between the spectrum of Rabi model and the parameters $\beta,\ N$. There are also have the sudden death and
sudden birth or revival of the entanglement for the two qubits in some proper regime for the proper parameters \cite{agar}.

\begin{widetext}
\begin{center}
\begin{figure}[ht]
\begin{tabular}{cc}
\includegraphics[width=0.4\textwidth]{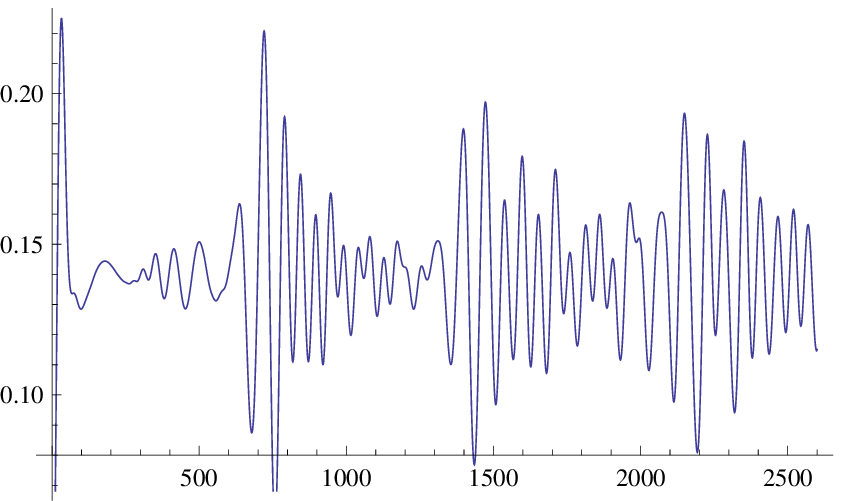}&
\includegraphics[width=0.4\textwidth]{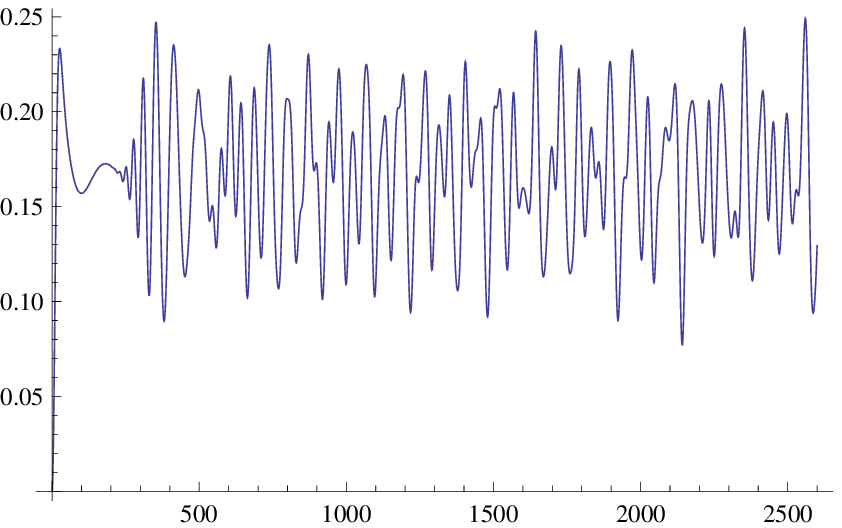}\\
\includegraphics[width=0.4\textwidth]{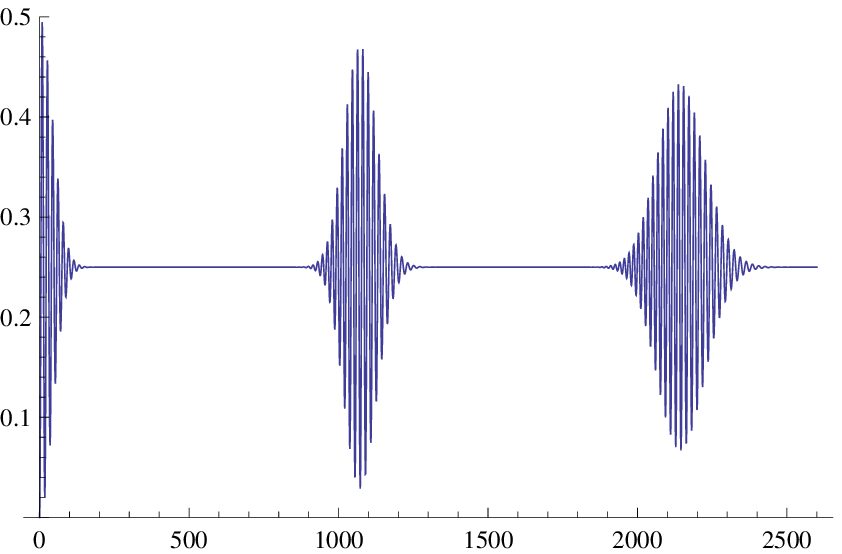}&
\includegraphics[width=0.4\textwidth]{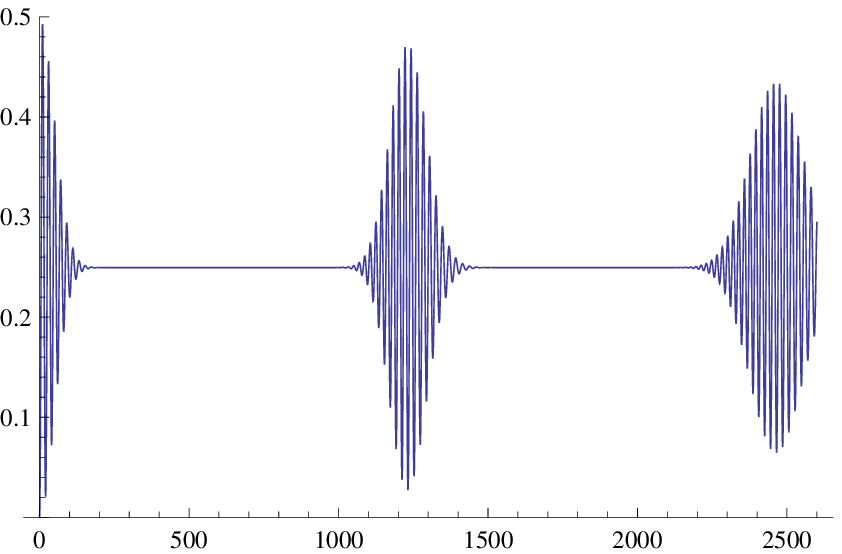}
\end{tabular}
\caption{Schematic diagrams of $T(t)$ with the four parameters as $|\alpha|^2=25,\ 16,\ \frac{\omega_0}{\omega}=0.23,\
\beta=0.26, \frac{\kappa}{\hbar\omega_0}=0.1$ for the top two figures and $|\alpha|^2=16,\ \frac{\omega_0}{\omega}=0.23,\ 0.2,\
\beta=0.12, \frac{\kappa}{\hbar\omega_0}=0.1$ to the bottom two figures respectively. The four figures ultrastrongly implies that
the parameter $a$ influences the qutrit dynamically and that the parameter $\frac{\omega_0}{\omega}$ has only very small influences
on the dynamics of $T(t)$ .}\label{fig1}
\end{figure}
\begin{figure}[ht]
\begin{tabular}{ccc}
\includegraphics[width=0.3\textwidth]{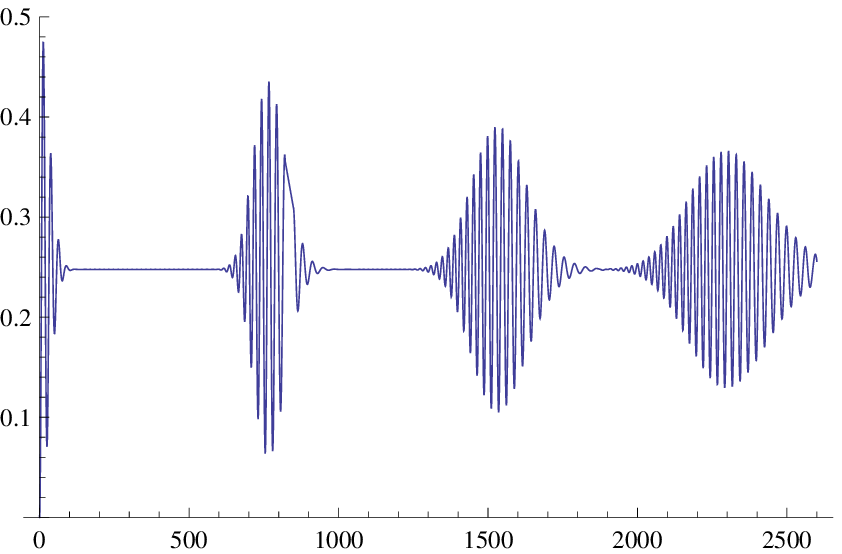}&
\includegraphics[width=0.3\textwidth]{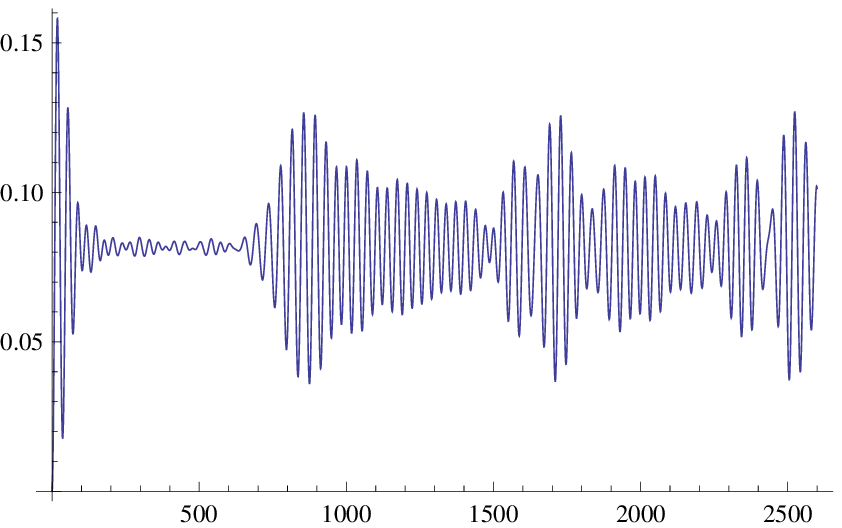}&
\includegraphics[width=0.3\textwidth]{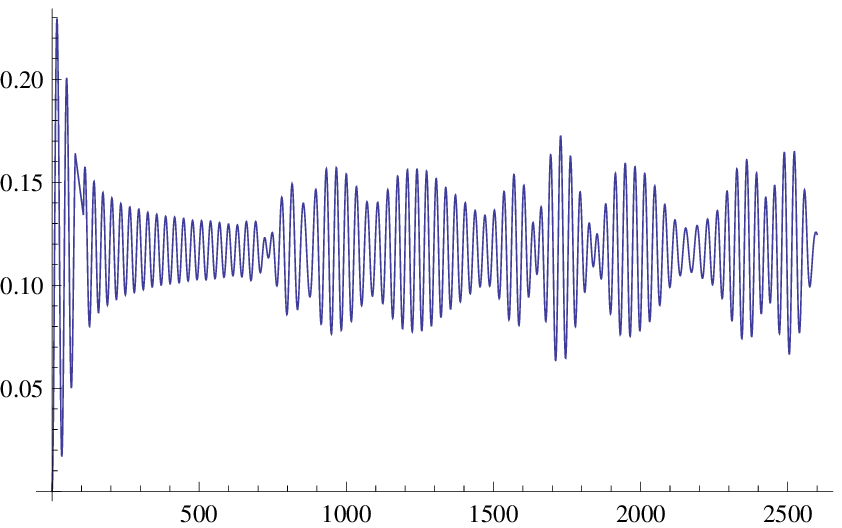}
\end{tabular}
\caption{Schematic diagrams of $T(t)$ with the same parameters  $\frac{\omega_0}{\omega}=0.2,\
\frac{\kappa}{\hbar\omega_0}=0.01$ and different parameters
$|\alpha|^2=16,\ \beta=0.16; \ |\alpha|^2=16;\ \beta=0.36,\ |\alpha|^2=20,\ \beta=0.36$,
The apparent difference in these three figures ultrastrongly implies that the parameters $\beta,\ |\alpha|^2$ influence the entanglement of the two qubits complicatedly.}\label{fig2}
\end{figure}
\end{center}
\end{widetext}

However, some anti-intuition phenomenal results might appear due to the unusual results (\ref{e0})-(\ref{epms}) where the coupling
parameter $\beta$ is nonlinearly and complicatedly involved \cite{iris}. So, we could take advantage of these intricate relations for entanglement maintaining. Choosing special parameters, we can make $T(t)$ very small, as the following figures show. From Fig.(\ref{fig3}), we see that $T(t)<0.015$ could be achieved by suitable choice of the parameters and the fully entangled state will have very high probability ( great than $97$ percents ) to remain unchanged. Three figures in Fig.(\ref{fig3}) ultrastrongly imply that the very small  change in parameters $\beta,\ |\alpha|^2$ does not influence the entanglement of the two qubits apparently.

\begin{widetext}
\begin{center}
\begin{figure}[ht]
\begin{tabular}{ccc}
\includegraphics[width=0.3\textwidth]{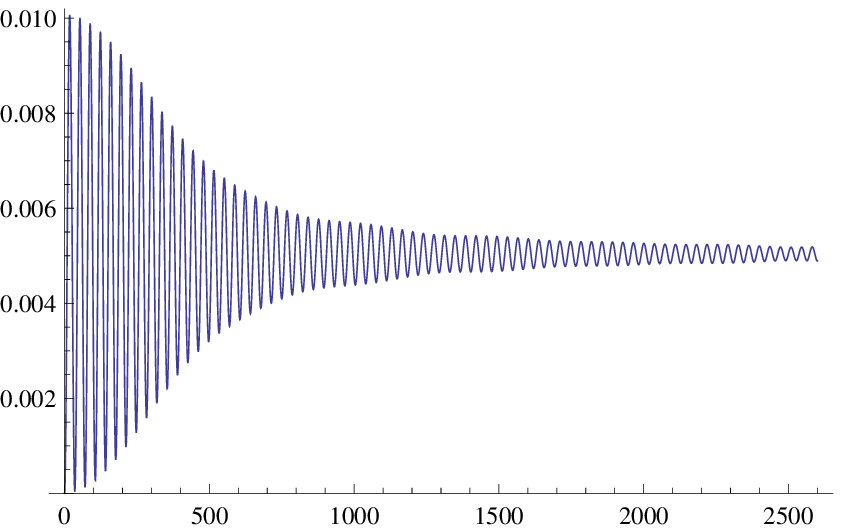}&
\includegraphics[width=0.3\textwidth]{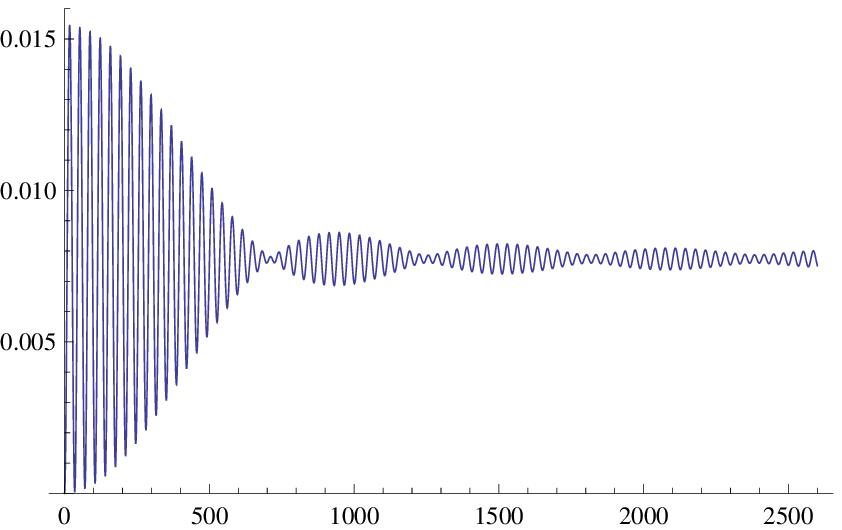}&
\includegraphics[width=0.3\textwidth]{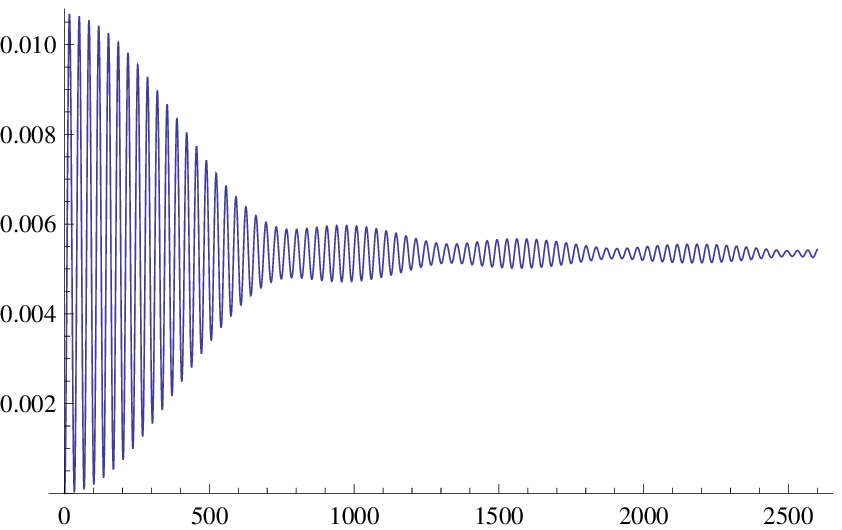}
\end{tabular}
\caption{Schematic diagrams of $T(t)$ with the same two parameters  $\frac{\omega_0}{\omega}=0.12,\ \frac{\kappa}{\hbar\omega_0}=0.02$ and the other different parameters $|\alpha|^2=106,\ \beta=0.4193; \ |\alpha|^2=116;\ \beta=0.4193,\ |\alpha|^2=106,\ \beta=0.4293,$ Three figures ultrastrongly imply that the very small change in parameters $\beta,\ |\alpha|^2$ does not influence the entanglement
of the two qubits apparently.}\label{fig3}
\end{figure}
\end{center}
\end{widetext}

\begin{figure}[ht]
\includegraphics[width=0.45\textwidth]{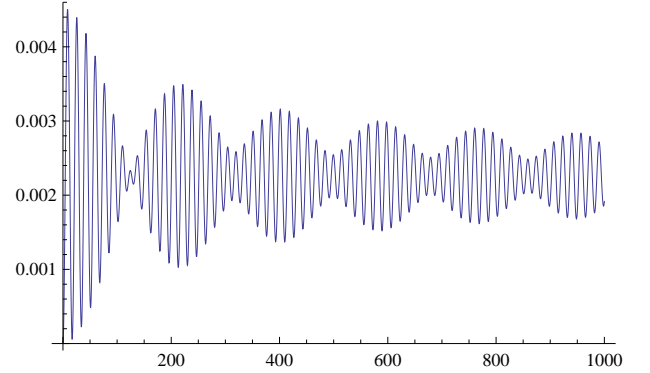}
\caption{$T(t)$ approaches zero as  the four parameters taking the following values
$\frac{\omega_0}{\omega}=0.2,\ \frac{\kappa}{\hbar\omega_0}=-0.7,\ |\alpha|^2=250,\ \beta=0.4717$. }\label{fig4}
\end{figure}

However, the further smaller $T(t)$ is not easy to obtain if the inter-qubit coupling parameter $\kappa$  is positive, which is
generally guaranteed in experiment.  We just suppose it is possible for $\kappa$ being negative for the reduction of $T(t)$. Numerical
result shows that it is possible to make $T(t)$ approach zero, see Fig(\ref{fig4}).
In this situation, it is reasonable to say that the two qubits could keep in its initial state
$\frac1{\sqrt2}(|\uparrow,\downarrow\rangle+|\downarrow,\uparrow\rangle)$ longer time. Here we do not consider the environmental influence
on the two qubits, as the ultro-strong coupling of qubits and oscillator is applied mainly in superconducting circuits, where
where charge noise has relatively little influence on some kinds of qibit, like the phase qubit, the capacitively shunted flux qubit,
and transmon qubit\cite{12011364,Joo,stef,koch}. Our study is applied to these circuit qubits. Numerical study shows that favorable
parameters for small $T(t)$ are $\beta $ ranging $0.3-0.5$, $|\alpha|^2$ far larger than $10$. Of course, negative interqubit
coupling will also make $T(t)$ easy approach zero. All these properties are shown somehow in Fig(\ref{fig1})-(\ref{fig4}).

We now compare our results with the general ones, which give death and revival phenomena. Various physical quantities in JC model
generally have the phenomena of death and rebirth, such as inversion, entanglement, etc. Here we give inverse $W(t)$ as in Ref.\cite{scu}
\begin{eqnarray}
W(t)=\sum_{N=0}^{\infty}p(N)\bigg[\frac{\Delta^2}{\bar{\Omega}_N^2}+\frac{4g^2(N+1)}{\bar{\Omega}_N^2}
\cos(\bar{\Omega}_Nt)\bigg],\label{wt}
\end{eqnarray}
where $\Delta$ is detuning parameter and $\bar{\Omega}_N=\Delta^2+4g(N+1)$ for the JC model with the coupling parameter $g$. These death and
rebirth phenomena occur because of the granular structure of photon distribution. As the physical quantities are the summation of the
revelent terms evolving with time as in Eq.(\ref{wt}). At the beginning, the revelent terms are correlated, and will become
un-correlated with time due to their related with different excitation (like $\Omega_N$ in Eq.(\ref{wt})) of JC Hamiltonian. So
the collapse happens. As time further goes, the restore occurs and revival happens too \cite{scu}.

Compared with Eq.(\ref{wt}), our formula for $T(t)$ is Eq.(\ref{t0}), where $\frac{\tilde{Y}_{N,+}^2}{\tilde{L}_{N,+}^4}$
 in every sum term is very complex and depends on the coupling strength $\beta$ and number $N$ nonlinearly.  Furthermore, it is possible for
$\frac{\tilde{Y}_{N,+}^2}{\tilde{L}_{N,+}^4}$  to be zero, which in contrast to the terms $\frac{4g^2(N+1)}{\Omega_N^2}$ being non-zero.
Because $p(N)$ is peaked at $|\alpha|^2=N$, $T(t)$ will approach zero if $\frac{\tilde{Y}_{N,+}^2}{\tilde{L}_{N,+}^4}$ as a function
of $N$ is extremely small in a relatively large interval around $N$ with a careful choice of the other parameters
$\frac{\omega_0}{\omega},\ \frac{\kappa}{\hbar\omega},\ \beta$. Due to the fact $\frac{\tilde{Y}_{N,+}^2}{\tilde{L}_{N,+}^4}$, both
$\kappa$ and $\Omega_{2N}$ negative will make $\frac{\tilde{Y}_{N,+}^2}{\tilde{L}_{N,+}^4}$ extremely small in
relative large interval with $\Omega^2_{1N}\rightarrow 0$ at $N=|\alpha|^2$.

In summary, coupled ultrastrongly with a quantum oscillator, the qubits with interqubit coupling dynamically evolved. The initial
fully entangled two qubits' state $\frac1{\sqrt2}(|\uparrow,\downarrow\rangle+|\downarrow,\uparrow\rangle)$ with the initial coherent state $|\alpha|^2$ for the quantum oscillator will exhibit some unusual characteristic by suitable choice of some parameters. The two qubits will have high possibility to be in its initial state $\frac1{\sqrt2}(|\uparrow,\downarrow\rangle+|\downarrow,\uparrow\rangle)
$ for some large $|\alpha|^2$ with the qubit-osillator' coupling ranging about in 0.3-0.5 and will have nearly to $1$ possibility
with the interqubit coupling negative. This unusual and novel result can only appear in the Rabi Hamiltonian system in strong coupling
regiem, where quantum eigenvalues complicatedly depends on the number $N$ and coupling strength $\beta$. They will provide new and
interesting properties for the phase qubit, the capacitively shunted flux qubit, and transmon qubit to utilized in quantum computation
and information process.

\acknowledgments
The work was partly supported by the National Natural Science of China (No. 10875018)
and the Major State Basic Research Development Program of China (973 Program: No.2010CB923202).

\end{document}